\documentclass[twocolumn,showpacs,amsmath,amssymb,prl,superscriptaddress,floatfix]{revtex4}

\usepackage{graphicx}
\usepackage{bm}
\usepackage{amsmath,amsfonts}

\begin{document}

\title{One-dimensional Rydberg Gas in a Magnetoelectric Trap}

\author{Michael Mayle}
\affiliation{Theoretische Chemie, Physikalisch--Chemisches Institut,
Universit\"at Heidelberg, Im Neuenheimer Feld 229, D-69120 Heidelberg,
Germany}

\author{Bernd Hezel}
\affiliation{Physikalisches Institut, Universit\"at Heidelberg,
Philosophenweg 12, D-69120 Heidelberg, Germany}

\author{Igor Lesanovsky}
\affiliation{Institut f\"ur Theoretische Physik, 
Universit\"at Innsbruck, A-6020 Innsbruck, Austria}

\author{Peter Schmelcher}
\affiliation{Theoretische Chemie, Physikalisch--Chemisches Institut,
Universit\"at Heidelberg, Im Neuenheimer Feld 229, D-69120 Heidelberg,
Germany}
\affiliation{Physikalisches Institut, Universit\"at Heidelberg,
Philosophenweg 12, D-69120 Heidelberg, Germany}

\date{\today}

\begin{abstract}\label{txt:abstract}
We study the quantum properties of Rydberg atoms in a magnetic
Ioffe-Pritchard trap which is superimposed by a homogeneous electric
field. Trapped Rydberg atoms can be created in long-lived electronic
states exhibiting a \emph{permanent} electric dipole moment of several
hundred Debye. The resulting dipole-dipole interaction in conjunction
with the radial confinement is demonstrated to give rise to an
effectively one-dimensional ultracold Rydberg gas with a
macroscopic interparticle distance. We derive analytical expressions for
the electric dipole moment and the required linear density of Rydberg
atoms.
\end{abstract}


\maketitle
Because of their widely tunable properties, ultracold atomic gases
provide the ideal playground to model and study complex many body
systems. The interatomic interaction can be tailored using
Feshbach-resonances and magnetic, optical, and electric fields can be
applied in order to generate virtually any external potential. Famous
examples for the versatility are the demonstration of the Mott-Insulator
to superfluid phase-transition of ultracold atoms in an optical lattice
\cite{Bloch02}, the BEC-BCS crossover in a gas of $^\text{6}$Li
\cite{Chin04}, or the Kosterlitz-Thouless phase transition studied
within a two-dimensional Bose-Einstein condensate \cite{Hadzibabic06}.
Traditionally, there is a great interest in studying systems with
reduced spatial dimensions, such as the latter example. One paradigm is
constituted by the work of Lieb and Liniger who were the first to solve
the system of pointlike interacting bosons in one dimension using
Bethe's ansatz \cite{Lieb63}. In the limit of an infinitely strong
interparticle interaction strength, a so-called Tonks-Girardeau gas
emerges \cite{GirardeauParedes}.

Besides gases of ground state atoms, particularly Rydberg gases
represent excellent systems to study the influence of a strong
interparticle interaction on the dynamics of many-particle systems. Due
to the large displacement of the ionic core and the valence electron,
Rydberg atoms can develop a large electric dipole moment leading to a
strong and long-ranged dipole-dipole interaction among them
\cite{Gallagher}. However, unlike ground state atoms Rydberg atoms
suffer from radiative decay and hence the mutual interaction time is
limited by the lifetime of the electronically excited state. But still,
in so-called frozen Rydberg gases, where the timescale of the atomic
motion is much longer than the radiative lifetime, exciting effects like
the dipole-blockade \cite{LukinTong} or resonant population transfer
\cite{PilletGallagher} have been observed.

Several works have focused on the important issue of trapping Rydberg
atoms based on electric \cite{Hyafil04}, optical \cite{Dutta00},
or strong magnetic fields \cite{Choi05_1}. Due to the high level density
and the strong spectral fluctuations with spatially varying fields,
trapping or manipulation in general is a delicate task. This holds
particularly for the case where both the center of mass and internal
motion are of quantum nature. Moreover, the inhomogeneous external
fields lead to an inherent coupling of these motions. Addressing this
regime, it has been recently theoretically shown that Rydberg atoms can
be tightly confined and prepared in long-lived electronic states in a
magnetic Ioffe-Pritchard (IP) trap \cite{Hezel} which can be
miniaturized using so-called atomchips \cite{FolmanFortagh}. Here we use
the IP configuration as a key ingredient in order to 'prepare' and study
a one-dimensional (1D) Rydberg gas. Specifically, we propose a modified IP trap, a
magneto-electric trap, which offers confining potential energy surfaces
for the atomic center of mass (c.m.)~motion in which the atoms possess
an oriented permanent electric dipole moment. We derive analytical
expressions for the dipole-dipole interaction among trapped Rydberg
atoms and estimate below which Rydberg atom density a 1D Rydberg gas is
expected to form. Moreover, we estimate the lifetime of such a gas.

Proceeding along the lines of Refs.~\cite{Hezel,Lesanovsky05_03}, we
employ a two-body approach in order to model an alkali metal atom in a
Rydberg state. We assume the single valence electron and the ionic core
(mass $M_c$) to interact via a pure Coulomb potential. While the
inclusion of the fine-structure and quantum defects can be readily done,
it turns out not to be necessary for high angular momentum electronic
states in the regime we are focussing on \cite{Hezel}. The IP field
configuration is given by
$\mathbf{B}(\mathbf{x})=B\mathbf{e}_3+\mathbf{B}_\text{lin}
(\mathbf{x})$
with
$\mathbf{B}_\text{lin}(\mathbf{x})=
G\left[x_1\mathbf{e}_1-x_2\mathbf{e}_2\right]$
and the vector potential reads
$\mathbf{A}(\mathbf{x})=
\mathbf{A}_\mathrm{c}(\mathbf{x})+\mathbf{A}_\mathrm{lin}(\mathbf{x})$
with
$\mathbf{A}_\mathrm{c}(\mathbf{x})=
\frac{B}{2}\left[x_1\mathbf{e}_2-x_2\mathbf{e}_1\right]$
and $\mathbf{A}_\mathrm{lin}(\mathbf{x})=Gx_1x_2\mathbf{e}_3$, where $B$
and $G$ are the Ioffe field strength and the gradient, respectively.
In addition, we apply a homogeneous electric field pointing in the
$x_1$-direction of the laboratory frame $\mathbf{F}=F\mathbf{e}_1$.
After introducing relative and c.m.\ coordinates ($\mathbf{r}$ and
$\mathbf{R}$) and employing the unitary transformation
$U=\exp\left[i\frac{B}{2}\mathbf{e}_3\times \mathbf{r} \cdot
\mathbf{R}\right]$, the Hamiltonian describing the
Rydberg atom becomes (atomic units are used unless stated otherwise)
\begin{eqnarray}
  H_\text{IPE}&=&H_A
  +\mathbf{A}_\text{c}(\mathbf{r})\cdot\mathbf{p}
  +\frac{\mathbf{\mathbf{P}}^2}{2M_c}
  -\mbox{\boldmath$\mu$}_N\cdot \mathbf{B}(\mathbf{R})+\mathbf{F}\cdot
    \mathbf{r}\nonumber\\&&
  -\mbox{\boldmath$\mu$}_e\cdot \mathbf{B}(\mathbf{R}
  +\mathbf{r})+\mathbf{A}_\text{lin}(\mathbf{R}+\mathbf{r})
	\cdot\mathbf{p
} .
\label{eq:hamiltonian_approximated}
\end{eqnarray}
Here, $H_A=\frac{\mathbf{p}^2}{2}-\frac{1}{r}$ is the Hamiltonian of
a hydrogen atom possessing the energies $E_{n}=-\frac{1}{2}n^{-2}$.
The second term denotes the energy of the electron in the
homogeneous Ioffe field due to its orbital motion. The following two
terms of $H_\text{IPE}$ describe the motion of a point-like particle
possessing the magnetic moment $\mbox{\boldmath$\mu$}_N$ in the
presence of the field $\mathbf{B}$. The magnetic moments are
connected to the electronic spin $\mathbf{S}$ and the nuclear spin
$\mathbf{\Sigma}$ according to $\mbox{\boldmath$\mu$}_e=-\mathbf{S}$
and $\mbox{\boldmath$\mu$}_N=-\frac{g_N}{2M_c}\mathbf{\Sigma}$, with
$g_N$ being the nuclear $g$-factor. We neglect the term involving
$\mbox{\boldmath$\mu$}_N$ in the following due to the large nuclear
mass. The electric field interaction, which in case of a neutral
two-body system couples only to the relative coordinates, gives rise to
the fifth term. The last two terms of $H_\text{IPE}$ are spin-field and
motionally induced terms coupling the electronic and c.m.~dynamics. We
focus on a parameter regime which allows us to neglect the diamagnetic
interactions \cite{Hezel}.

In order to find the stationary states of the Hamiltonian
(\ref{eq:hamiltonian_approximated}), we assume that neither the
magnetic nor the electric field causes couplings between electronic
states with different principal quantum number $n$. In this case we
can consider each $n$-manifold separately and may represent the
Hamiltonian (\ref{eq:hamiltonian_approximated}) in the space of the
$2n^2$ states which span the $n$-manifold under investigation. The
parameter range in which this approximation is valid has been
thoroughly discussed in Refs.~\cite{Hezel,Schmidt07}. Because of the
translational symmetry of the IP and the electric field, the axial
c.m.~motion along $Z$ can be separated from the transversal motion
in the $X$-$Y$ plane. If we omit the energy offset $E_n$ and
introduce scaled c.m.~coordinates ($\mathbf{R}\rightarrow
\gamma^{-\frac{1}{3}} \mathbf{R}$ with $\gamma=G M_c$) while scaling
the energy unit with $\epsilon_\mathrm{scale}=\gamma^\frac{2}{3}/M_c$,
we arrive at the Hamiltonian
\begin{eqnarray}
  \mathcal{H}=\frac{P_1^2+P_2^2}{2}+\mbox{\boldmath$\mu$}\cdot
\mathbf{G}+\gamma^{-\frac{2}{3}}M_c\left[\mathcal{H}_m+\mathcal{H}
_e\right].
  \label{eq:working_hamiltonian}
\end{eqnarray}
This Hamiltonian governs the transversal c.m.\ as well as the
electronic dynamics and involves the effective magnetic field
$\mathbf{G}=
X\mathbf{e}_1-Y\mathbf{e}_2+\gamma^{-\frac{2}{3}}M_cB\mathbf{e}_3$.
The symbols $\mbox{\boldmath$\mu$}$, $\mathcal{H}_m$, and
$\mathcal{H}_e$ are the $2n^2$-dimensional matrix representations of
the operators $\frac{1}{2}\left[\mathbf{L}+2\mathbf{S}\right]$,
$H_m=\mathbf{A}_\mathrm{lin}(\mathbf{r})\cdot\mathbf{p}+
\mathbf{B}_\mathrm{lin}(\mathbf{r})\cdot\mathbf{S}$, and the electric
field interaction $H_e= F x$, respectively (we introduced
$\mathbf{L}=\mathbf{r}\times \mathbf{p}$). A thorough
interpretation of this Hamiltonian in the absence of
$\mathcal{H}_e$ is provided in Ref.~\cite{Hezel}. In order
to solve the corresponding Schr\"odinger equation we employ an
adiabatic approach. To this end an unitary transformation $U(X,Y)$
which diagonalizes the last two (matrix) terms of the Hamiltonian is applied, 
$U^\dagger(X,Y)(\mbox{\boldmath$\mu$}\cdot\mathbf{G}+
\gamma^{-\frac{2}{3}}M_c\left[\mathcal{H}_m+
\mathcal{H}_e\right])U(X,Y)=E_\alpha(X,Y)$. 
Since $U(X,Y)$ depends on the c.m.\ coordinates, the transformed kinetic
energy term involves non-adiabatic (off-diagonal) coupling terms which 
can be neglected in our parameter regime \cite{Hezel}. We are thereby
led to a set of $2n^2$ decoupled differential equations governing the
adiabatic c.m.\ motion within the individual two-dimensional energy
surfaces $E_\alpha(X,Y)$, i.e., the surfaces $E_\alpha(X,Y)$ serve as
potentials for the c.m.\ motion of the atom.
\begin{figure}\center
\includegraphics[scale=0.65]{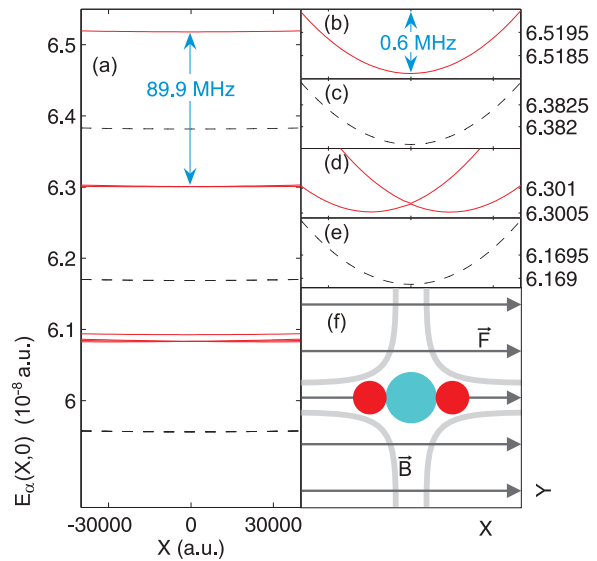}
\caption{(Color online) Potential energy surfaces of the c.m.\ motion of
a $^{87}$Rb atom ($n=30$) in an IP trap with
$B=10\,\text{G}$,
$G=10\,\text{Tm}^{-1}$.
Dashed lines:
$F=0$, solid lines:
$F=5.14\,\text{Vm}^{-1}$.
An overview of the seven energetically highest potential curves is
shown in panel (a). Magnified views of the uppermost (b,c) and next
lower ones (d,e) are also provided. The range of the $X$-coordinate,
corresponding to $2.1\mu\text{m}$, is the same for each subfigure
(a)-(e). The total field configuration is sketched in panel (f) where
the circles depict the locations of the minima of the uppermost (big
circle) and the two adjacent lower-lying (small circles) c.m.\ surfaces.
The magnetic field lines are indicated in gray while the electric field
is sketched by black arrows.} \label{fig:surfaces}
\end{figure}
In Fig.~\ref{fig:surfaces} we present intersections along the
$X$-direction of such potential surfaces for $B=10\,\text{G}$,
$G=10\,\text{Tm}^{-1}$ and  $n=30$ in the case of $^{87}$Rb. For
zero electric field strength (dashed lines), the potential curves are
organized in groups which are energetically well-separated by a gap of
$\gamma^{-\frac{2}{3}}M_cB=89.9$ MHz. The uppermost surface is
non-degenerate and provides an approximately harmonic confinement with a
trap frequency of $\omega=G\sqrt{n/2BM_c}=13.9\,$kHz corresponding to
$0.1\,\mu\text{K}$. The two adjacent lower surfaces are degenerate and
also approximately harmonic. As soon as an electric field is applied,
all surfaces are shifted considerably in energy. This is visible from
the solid curves in Fig.~\ref{fig:surfaces} for which an electric field
of strength $F=5.14\,\text{Vm}^{-1}$ is applied. The shapes of the
potentials are barely affected by the electric field such that Rydberg
states which were trapped in a pure IP configuration remain confined
also in the magneto-electric trap. Moreover, adding the electric field
leads to non-trivial effects: The second and third surface, which were
almost degenerate in the absence of the electric field, are now shifted
in opposite ways along the $x_1$-direction. All surfaces shown provide a
harmonic confinement with a trap frequency $\omega$ also in the
$x_2$-direction. We remark that the chosen parameter set does not
generate an extreme constellation hence an even stronger confinement can
be achieved without invalidating the applied approximations
\cite{Hezel}.

\begin{figure}
\includegraphics[scale=1]{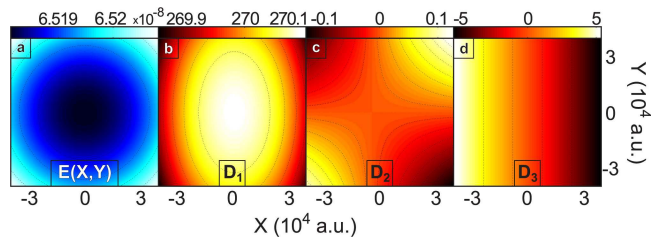}
\caption{(Color online) (a) Uppermost electronic potential surface
for the c.m.\ motion of $^{87}$Rb in the $n=30$ multiplet and
the parameters used in Fig.~\ref{fig:surfaces}. (b-d) Components of
the electronic dipole moment $\mathbf{D}(\mathbf{R})$ in atomic units.
One recognizes the clear alignment of the electric dipole moment along
the electric field vector. 
The numerically calculated values of $\mathbf{D}(\mathbf{R})$ are to
good accuracy reproduced by Eq.~(\ref{eq:dipole_moment}).}
\label{fig:uppermost_surface}
\end{figure}

Let us now investigate the electronic properties of a Rydberg atom
being trapped in the uppermost potential surface. For $F=0$ and
sufficiently large values of $B$, this surface is formed almost
exclusively by the highest possible electronic angular momentum
state, i.e., $l=n-1$ \cite{Hezel}. If $F$ is increased, electronic
states with smaller $l$ will be inevitably admixed to the electronic
state belonging to this energy surface. An interesting property to
investigate is hence the electric dipole moment of trapped Rydberg
atoms: While for $F=0$ the electronic states are almost pure parity
eigenstates and therefore exhibit almost no electric dipole, the
admixture of lower $l$ states to the uppermost surface in the presence 
of the field is expected to give rise to a non-vanishing expectation
value of the dipole operator. Indeed, this becomes evident in
Fig.~\ref{fig:uppermost_surface} where the uppermost potential surface
and the three components of the expectation value of the electric dipole
operator $\mathbf{D}(\mathbf{R})=\left<\mathbf{r}\right>(\mathbf{R})$
are shown (same parameters as in Fig.~\ref{fig:surfaces}). It can
be clearly seen that a permanent dipole moment is established whose
dominant contribution points along the electric field vector.

In order to study the dependence of $\mathbf{D}(\mathbf{R})$ on the
field strengths $F$ and $B$ as well as on the degree of electronic
excitation, we use perturbation theory. In the limit of a large
Ioffe field strength $B$ the unitary transformation which
diagonalizes the Hamiltonian (\ref{eq:working_hamiltonian}) can be
written explicitly as $U_\mathbf{r}=e^{-i \alpha (L_x+S_x)}e^{-i \beta
(L_y+S_y)}$ with $\sin\alpha = -Y\,|\mathbf{G}|^{-1}$,
$\cos\alpha=\sqrt{|\mathbf{G}|^{2}-Y^2}\,|\mathbf{G}|^{-1}$,
$\sin\beta = X\,(|\mathbf{G}|^2-Y^2)^{-\frac{1}{2}}$ and $\cos\beta =
\gamma^{-\frac{2}{3}}M_cB\,(|\mathbf{G}|^2-Y^2)^{-\frac{1}{2}}$. 
This transformation rotates the $z$-axis into the local magnetic field
direction where $\alpha$ and $\beta$ denote the rotation angles. Using
this result we find up to first order in $F/B$ the electric dipole
moment (in atomic units)
\begin{eqnarray}
  \mathbf{D}(\mathbf{R})=\frac{9}{2}\frac{F}{B}n^2(n-1)\left(
  \begin{array}{c}
  \cos\beta \\
  \sin\beta\sin\alpha \\
  \sin\beta\cos\alpha \\
  \end{array}
\right)\label{eq:dipole_moment}.
\end{eqnarray}
We note that $\mathbf{D}(\mathbf{R})$ scales proportional to the third
power of the principal quantum number and can therefore gain a
significant magnitude even if the ratio $F/B$ is small. Good agreement
of Eq.~(\ref{eq:dipole_moment}) with the calculated data presented in
Fig.~\ref{fig:uppermost_surface} is found; e.g., in the vicinity of the
minimum of the potential surface ($X=Y=0$) we find an exact value of
$D_x=270$ whereas the expression (\ref{eq:dipole_moment}) yields $276$.
For the remaining components Eq.~(\ref{eq:dipole_moment}) yields zero at
the origin. For smaller ratios of $F/B$, even better agreement can be
achieved.

Because of the dependence on the angles $\alpha$ and $\beta$, the dipole
moment depends weakly on the quantum state of the c.m.~motion. However,
since the field configuration is translationally symmetric the electric
dipole moment is independent of the $Z$-position of the Rydberg atoms in
the trap. If we now consider two transversally confined atoms in the
same trap at the longitudinal positions $Z_A$ and $Z_B$, we can write
for their dipole-dipole interaction
\begin{eqnarray}
V_\mathrm{D}(\mathbf{R}_A,\mathbf{R}_B)
=\frac{1}{|\mathbf{R}_A-\mathbf{R}_B|^3}
\Big(\mathbf{D}(\mathbf{R}_A)\cdot\mathbf{D}(\mathbf{R}
_B)\nonumber\\
-3\left[\mathbf{D}(\mathbf{R}_A)\cdot
\mathbf{e}\right]\left[\mathbf{D}(\mathbf{R}_B)\cdot
\mathbf{e}\right]\Big)\approx\frac{\mathbf{D}(\mathbf{R}_A)\cdot\mathbf{
D}
(\mathbf{R}_B)}{|Z_A-Z_B|^3}\label{eq:dipole_potential}
\end{eqnarray}
where $\mathbf{e}$ denotes the interparticle unit vector. The
approximation in Eq.~(\ref{eq:dipole_potential}) holds due to the
orientation of the dipoles and the assumption that $|Z_A-Z_B|$ is large
compared to the transversal oscillator length of the trap. These
conditions moreover ensure a minimal coupling of the transversal and
longitudinal motion. Using this approximation one can estimate the
interaction energy of one atom being part of an infinite atomic chain
with an interparticle spacing $a$. One finds
\begin{eqnarray} 
E_\mathrm{int}=2\frac{\mathbf{D}^2(0)}{a^3}\sum_{k=1}^\infty k^{-3}=
\frac {81}{2a^3}\frac{F^2}{B^2}n^4(n-1)^2\zeta(3)
\end{eqnarray}
with the Riemann zeta function $\zeta(x)$. Here we have approximated
$\mathbf{D}^2(\mathbf{R})\approx \mathbf{D}^2(0)$ since the dipole
moment barely varies in the vicinity of $X=Y=0$. If the interaction
energy $E_\mathrm{int}$ is smaller than the transversal trap frequency
$\omega$, we can assume that the interacting atoms remain in the
transversal ground state: This is considered the 1D regime. The linear
density below which a 1D Rydberg gas is expected to form is then given
by
\begin{equation}
N_\mathrm{1D}=\frac{\sqrt{B}}{3}\left[3\sqrt{\frac{M_c}{2}}\frac{F^2}{G}
\, \zeta(3)\, n^{7/2}(n-1)^2\right]^{-\frac{1}{3}}\;.
\label{eq:density}
\end{equation}
Above this density, excited transversal c.m.\ states might be populated
resulting in a quasi 1D Rydberg gas which is certainly of interest on
its own. For our parameter set, we obtain a minimal interparticle
spacing $a=43\,\mu$m; hence a chain of 1 mm in length contains 23
particles. This density can be further increased by either increasing
the magnetic field gradient and/or decreasing the electric field
strength: At $B=10\,$G, $G=100\,\text{Tm}^{-1}$, and
$F=0.514\,\text{Vm}^{-1}$ a chain of the same length would contain 230
Rydberg atoms.

Finally, the issue of radiative decay has to be addressed. Since
the electric field admixes merely a few $l=m$ states ($l<n-1$) to
the electronic wave function of the uppermost surface, its circular
character remains dominant resulting in one prevalent decay channel. 
For an atom being confined to the energy surface which is shown in
Fig.~\ref{fig:uppermost_surface}, we have calculated a lifetime of
$\tau\approx 2.1\,$ms which is in good agreement with the corresponding
field-free result $\tau(n,n-1)\approx
\frac{3}{2c^2}\left(\frac{n}{\alpha}\right)^5$ \cite{Jentschura05}.
Corrections to this bare decay rate are found to be of the order of
$(F/B)^2n^3$. Due to the scaling proportional to $n^5$, the lifetime can
be significantly enhanced by exciting to a higher principal quantum
number $n$. In addition, it can be further prolonged by establishing an
adapted experimental setup which inhibits the electromagnetic field mode
at the dominant transition frequency \cite{Hulet85}. At the same time, a
cryogenic environment will diminish the undesirable effect of stimulated
(de-)excitation by blackbody radiation. The timescale of the dynamics of
the Rydberg chain on the other hand depends on the field strengths via
the dipole moment and the interparticle spacing: A harmonic
approximation of the dipole-dipole interaction yields the one-particle
oscillator frequency
$\omega_\text{dd}=\sqrt{24\,\mathbf{D}^2(0)/M_ca^5}$.
As an example, the field configuration
$B=10\,$G, $G=100\,\text{Tm}^{-1}$, and $F=0.514\,\text{Vm}^{-1}$ yields
a timescale of less than 1\,ms.

Let us now briefly comment on the realization of such a Rydberg gas
which is certainly a challenging experimental task. One could start from
an extremely dilute ultracold atomic gas prepared in an elongated IP
trap. For transferring ground state atoms to high angular momentum
Rydberg states, techniques such as crossed electric and magnetic
fields or rotating microwave fields can be employed, see
Ref.~\cite{Lutwak97} and references therein. For low angular momentum
states, trapping and the formation of a permanent dipole represent still
an open question since quantum defects, spin-orbit coupling and reduced
radiative lifetimes have to be taken into account. During the
preparation, the excitation lasers have to be focussed such that Rydberg
atoms emerge only at positions separated by the interparticle spacing
$a$ which is required to meet the criterion (\ref{eq:density}). Since
$a$ is in the order of several $\mu$m, which can be resolved optically,
this should be feasible. The large value of $a$ moreover ensures that
the mutual ionization due to the overlap of the electronic clouds of two
atoms does not occur. For our circular states with $n=30$, the atomic
extension can be estimated by $\langle r\rangle\approx n^2=48$ nm and is
thus orders of magnitude smaller than the corresponding value of $a$ for
our field configuration. In order to probe the dynamics of the resultant
Rydberg chain, one can field-ionize the atoms: From the spatially
resolved electron signal a direct mapping to the positions of the
Rydberg atoms should be possible.

In conclusion, we demonstrated that in a magneto-electric trap Rydberg
states can be confined in electronic states exhibiting a permanent
electric dipole moment of hundreds of Debyes. Analytical expressions for
the density which is required to enter the 1D regime were calculated.
Moreover, we pointed out that the lifetime of the Rydberg states is
sufficiently long to probe the dynamics of the interacting gas. This
regime is complementary to the well-studied frozen Rydberg gases where
mechanical atom-atom interaction effects can hardly be probed. The
potential of the proposed magneto-electric trap is by far not entirely
exhausted; e.g., one could think of using the double-well structure
which is visible in Fig.~\ref{fig:surfaces}(d) in order to realize two
coupled dipolar Rydberg chains.

This work was supported by the German Research Foundation (DFG) under
the contract SCHM\,885/10-2 and within the framework of the
Excellence Initiative through the Heidelberg Graduate School of
Fundamental Physics (GSC~129/1). M.M.\ acknowledges support from the
Landesgraduier\-ten\-f\"orderung Baden-W\"urttemberg.


\vspace{-0.4cm}

\begin{thebibliography}{40}
\bibitem{Bloch02} M. Greiner \textit{et al.},  Nature \textbf{415},
39 (2002)
\bibitem{Chin04} C. Chin \textit{et al.}, Science \textbf{305}, 1128
(2004)
\bibitem{Hadzibabic06} Z. Hadzibabic \textit{et al.}, Nature
\textbf{441}, 1118 (2006)
\bibitem{Lieb63} E. H. Lieb and W. Liniger, Phys. Rev. \textbf{130},
1605 (1963); E. H. Lieb, \textit{ibid.} \textbf{130}, 1616 (1963)
\bibitem{GirardeauParedes} M. Girardeau, J. Math. Phys. \textbf{1}, 516
(1960); B. Paredes \textit{et al.}, Nature \textbf{429}, 277 (2004)
\bibitem{Gallagher} T.F. Gallagher, \textit{Rydberg Atoms}, Cambridge
University Press 1994
\bibitem{LukinTong} M. D. Lukin {\it{et al.}}, Phys. Rev. Lett.
\textbf{87}, 037901 (2001); D. Tong {\it{et al.}}, \textit{ibid.}
\textbf{93}, 063001 (2004)
\bibitem{PilletGallagher} W. R. Anderson, J. R. Veale, and T. F.
Gallagher Phys. Rev. Lett. \textbf{80}, 249 (1998); I. Mourachko
\textit{et al.}, \textit{ibid.} \textbf{80}, 253 (1998)
\bibitem{Hyafil04} P. Hyafil {\it{et al.}}, Phys. Rev. Lett.
\textbf{93}, 103001 (2004)
\bibitem{Dutta00} S.K. Dutta {\it{et al.}}, Phys. Rev. Lett.
\textbf{85}, 5551 (2000)
\bibitem{Choi05_1} J.-H. Choi {\it{et al.}}, Phys. Rev. Lett.
\textbf{95}, 243001 (2005)
\bibitem{Hezel} B. Hezel, I. Lesanovsky, and P. Schmelcher, Phys. Rev.
Lett. \textbf{97}, 223001 (2006); arXiv:0705.1299v2
\bibitem{FolmanFortagh} J. Fortagh and C. Zimmermann, Rev. Mod. Phys.
\textbf{79}, 235 (2007); R. Folman \textit{et al.}, Adv. At. Mol. Opt.
Phys. \textbf{48}, 263 (2002)
\bibitem{Lesanovsky05_03} I. Lesanovsky and P. Schmelcher, Phys. Rev.
Lett. \textbf{95}, 053001 (2005)
\bibitem{Schmidt07} U. Schmidt, I. Lesanovsky, and P. Schmelcher, J.
Phys. B \textbf{40}, 1003 (2007)
\bibitem{Jentschura05} U. D. Jentschura {\it{et al.}}, J. Phys. B
\textbf{38}, S97 (2005)
\bibitem{Hulet85} R. G. Hulet, E. S. Hilfer, and D. Kleppner, Phys. Rev.
Lett. \textbf{55}, 2137 (1985)
\bibitem{Lutwak97} R. Lutwak {\it{et al.}}, Phys. Rev. A \textbf{56},
1443 (1997)
\end{thebibliography}
\end{document}